\documentclass{article}

\usepackage[letterpaper,hmargin=1in,vmargin=1.25in]{geometry}
\usepackage[utf8]{inputenc}
\usepackage{amsmath}
\usepackage{bbding}
\usepackage{url}
\usepackage{booktabs}
\usepackage[english]{babel}
\usepackage{diagbox}
\usepackage{colortbl}
\usepackage{listings}
\usepackage{multirow}
\usepackage{xspace,xcolor}
\usepackage{tikz}
\usepackage{wrapfig}
\usepackage{caption,subcaption}
\usepackage{relsize}
\usepackage{booktabs}
\usepackage{amsmath,amsfonts,amssymb}
\usetikzlibrary{plotmarks}

\definecolor{bluemod}{rgb}{0.03,0.27,0.51}
\definecolor{light-gray}{gray}{0.989}
\definecolor{colKeys}{rgb}{0,0,1}
\definecolor{orange}{rgb}{1,0.5,0}
\definecolor{colIdentifier}{rgb}{1,0.5,0}
\definecolor{colString}{rgb}{0.5,0.1,0.1}

\newcommand{\cref}[1]{\cellcolor{green!20}\hspace*{-0.1cm}\texttt{#1}\hspace*{-0.1cm}}

\newcommand{\cini}[1]{\cellcolor{red!20}#1}
\newcommand{\crst}[1]{\cellcolor{gray!20}#1}

\newcommand{\bref}{\cellcolor{red!30}$\curvearrowleft$}
\newcommand{\bxef}{\cellcolor{gray!20}$\times$}
\newcommand{\mrow}[2]{\multirow{#1}{*}{#2}}

\newcommand{\smartcard}[2]{#1martcard#2\xspace}

\title{\textbf{High Precision Fault Injections on the Instruction Cache of ARMv7-M Architectures}}
\date{}

\usepackage{authblk}
\author[1,2,4]{Lionel Rivière\thanks{This work was partially funded by the French ANR E-MATA HARI Project}}
\author[1]{Zakaria Najm}
\author[1]{Pablo Rauzy}
\author[1,3]{Jean-Luc Danger}
\author[2]{Julien Bringer}
\author[1,3]{Laurent Sauvage}
\affil[1]{Institut Mines-Télécom ; Télécom ParisTech ; CNRS LTCI \\
  \{\textit{firstname}.\textit{lastname}\}@telecom-paristech.fr}
\affil[2]{SAFRAN Morpho \\
  \{\textit{firstname}.\textit{lastname}\}@morpho.com}
\affil[3]{Secure-IC \\
  \{\textit{firstname}.\textit{lastname}\}@secure-ic.com}
\affil[4]{Identity \& Security Alliance (The Morpho and Télécom ParisTech Research Center)}

\begin{document}
\maketitle
\begin{abstract}
%commentaires zak :
%Je pense qu'il faut dire que l'on fait une caractérisation complète du modèle comportemental du microcontrôleur
	%Lorque le mecanisme du cache est visé et soumis à une impultion electromegnétique (transient fault)
%Zak : a mon avie il vaux peut etre mieux de ne pas citer le nom du microcontroleur
% car on ne vise pas ce microcontroleur en particulier mais la famille de micro à base de
% CPU de la famille ARMV7-ME (donc parler de uC à bas
Hardware and software of secured embedded systems are prone to physical attacks.
In particular, fault injection attacks revealed vulnerabilities on the data and the control flow
allowing an attacker to break cryptographic or secured algorithms implementations.
While many research studies concentrated on successful attacks on the data flow, only a few targets the instruction flow.
In this paper, we focus on electromagnetic fault injection (EMFI) on the control flow,
especially on the instruction cache.
We target the very widespread (smartphones, tablets, settop-boxes, health-industry monitors and sensors, etc.) ARMv7-M architecture.
We describe a practical EMFI platform and present a methodology providing high control level and high reproducibility over fault injections.
Indeed, we observe that a precise fault model occurs in up to 96\% of the cases.
We then characterize and exhibit this practical fault model on the cache that is not yet considered in the literature.
We comprehensively describe its effects and show how it can be used to reproduce well known fault attacks.
Finally, we describe how it can benefits attackers to mount new powerful attacks or simplify existing ones.
\end{abstract}


%*************************************
%	INTRODUCTION
%*************************************
\section{Introduction}

Secured embedded systems, such as \smartcard{s}{s} or secure elements, store personal and sensitive data.
Individuals seamlessly rely on those physical objects for identity, banking, or telecommunication services for instance.
Although cryptographic and secure algorithms are mathematically strong,
their hardware and software implementations tend to be vulnerable to physical attacks.
In particular, fault injection attacks target physical implementations of secured devices in order to induce exploitable errors.
In this paper, we focus on electromagnetic fault injection (EMFI) on the instruction cache memory of a widespread hardware architecture.
%% PR: je répète EMFI parce que l'intro et la conclusion doivent pouvoir se lire facilement indépendemment du reste, comme l'abstract, donc ne pas dépendre d'accronymes introduit ailleurs dans le papier.
%We start by describing how such memories are implemented in modern architectures and why it is of a particular interest for an attacker.

\medskip
%***************************************************************************************
\subsection{Memory in Modern Embedded Architecture}

Modern microcontrollers use fast computing units processing at the nano-second time scale.
This makes it possible, for a given program, to perform hundreds of millions of operations in less than a second.
The program to be executed consists in data and instructions that are stored in non-volatile memories (Flash NVRAM, ROM).
Therefore, memory modules must be fast enough to provide such bandwidth to the CPU and avoid bottlenecks.
However, memories vary in size (in byte) and speed (response time and bandwidth), and their cost increase proportionally.
Thus, it is rarely possible to hold a whole program in a fast memory due to size limitations.
In order to provide a sufficient amount of fast memory at a fair price,
the industry has adopted a hierarchical memory structure providing cache mechanisms.

\medskip
%* * * * * * * * * * * * * * * * * * * * * * * * * * * * * * * * * * * * * * * * * * * *
\subsubsection*{Cache Mechanisms}
We here define a few concepts that are useful to get a better understanding of caches.\\
\emph{Locality principle:} instructions that are close (in the code) to the currently executed one are more likely to be executed soon (e.g., the next instructions, and the previous ones when in loops).
This is also true for data accessed from memory.\\
\emph{Temporal locality:} recently accessed items are likely to be accessed again in the near future (again, think of loops).\\
\emph{Spatial locality:} items that are close in term of memory location (address) are likely to be referred together (think of arrays, or big numbers such as cryptographic keys).\\
\emph{Cache miss:} a failed attempt to read data or an instruction from the cache because it is not there, which results in a main memory access with much longer latency.\\
\emph{Cache penalty:} the delay induced by a necessary access to the main memory in case of cache misses.

On ARM Cortex-M4 microcontrollers, which are members of the ARMv7-M family, the Least Recently Used (LRU) policy is used to determine which lines to replace in the instruction memory cache.

\medskip
%* * * * * * * * * * * * * * * * * * * * * * * * * * * * * * * * * * * * * * * * * * * *
\subsubsection*{Modern Microcontroller Design}

Nowadays, efficient microcontroller architectures rely on one or more cache levels.
Usually, level one cache memory, denoted as L1, is the closest to the CPU.
It consists of super fast memories but is of limited size (up to 64 kBytes on the ARM Cortex-M4).
In addition, the caching mechanism usually contains an additional instruction prefetch buffer queue.
It allows to mitigate delays from a change-of-flow in the instructions due to branching, subroutine calls, and possibly even system calls or interrupts.

\subsection{Contributions}

Our first contribution is to enhance the precision of practical electromagnetic fault injection attacks.
To this end, we expose the methodology that allowed us to obtain high control level and high reproducibility over fault injections.
A behavioral analysis on targeted programs makes it possible to extract the underlying fault model and to tune the experimental settings to increase the fault occurrence rate accordingly.

Our second contribution is a comprehensive interpretation of our EMFI effects on the cache mechanism of a modern micro\-controller without anti-glitch countermeasures enabled.
It reveals a high occurrence, predictable and easily reproducible fault model that can be described using two commonly accepted models: the instruction replacement and the instruction skip.
We exhibit how tampering with the instruction cache can lead to powerful attacks on the control flow.

\subsection{Organization of the Paper}
In Section~\ref{sec:FA} we explain why implementations are prone to physical fault injection and we review some existing attacks in the literature.
In Section~\ref{sec:cacheAttack} we expose the cache mechanism of our target device and then present in details how we performed EMFI on it.
In Section~\ref{sec:cacheEMFI} we depict our experimental setup and the target configuration settings.
We provide the experimental protocols that we used to perform the fault injections along with the corresponding observed behaviors as well as a cost assessment as required in the Common Criteria~\cite{cc}.
Section~\ref{sec:results} explains our results in more details, thanks to an extensive study of an assembly test code.
We then review the capabilities of the exhibited fault model with respect to existing attacks and potential new ones, including against existing countermeasures, in Section~\ref{sec:discuss}.

%***************************************************************************************
\section{Physical Attacks}
\label{sec:FA}

The security of most cryptographic primitives is mathematically proven.
However, the research community continually exposes new attacks,
including ones that are not the result of classical cryptanalysis, since the 90s.
In these cases, cryptographic algorithms are not directly defeated but flaws appear in their implementation.
Indeed, whether passive~\cite{DBLP:conf/crypto/KocherJJ99}, active~\cite{boneh-fault} or combined, physical attacks breaks cryptographic and secured implementations on embedded systems.

Any algorithm implementation is transcribed into a low-level machine language in order to be processed by a hardware device.
The compilation process, which is (hopefully) semantic preserving, associates data to operands and instructions to operators.
In the end, both data and instructions are stored in a non-volatile memory.
They will be moved from different memories (RAM, caches, and registers) via buses before being processed by the CPU.
Cryptographic primitives and software security features are of course no exception and follow this scheme.
Consequently, there is a strong necessity to understand the underlying mechanisms of the hardware layer in order to avoid exposing sensitive data to physical attacks.
%% PR: très joli paragraphe.

%***************************************************************************************
%\subsection{Cache Attacks in the Literature}
\medskip
%Timing Attacks~\cite{kocher-timing_attacks} (TA) are a sub-class of
Side-Channel Attacks~\cite{DBLP:conf/crypto/KocherJJ99} (SCA) exploit the dependency between the operation duration of an algorithm and the value being processed. %exemple du verifyPIN ?
These attacks can be straightforward when the number of executed instructions depends on the secret key: for instance the number of executed instruction in the square-and-multiply exponentiation algorithm depends linearly in the number of 1s bits in the key if no countermeasures are implemented.
In other cases, the time necessary to execute an operation will vary depending on the presence of the required data in cache memories.
The principle of attacks in such cases is simple and consists in exploiting cache penalties induced by cache misses that could occur during cryptographic computation.
Moreover, an active attacker could influence this parameter by faulting the cache address memory.

%% \textbf{[SCA]} "Cache timing attacks on the instruction path is harder to put into practice than data cache timing attacks"
%% PR: bien vu cette quote, j'ai trouvé une source et je l'ai intégré au paragraphe ci-dessous.

Most of existing fault attacks on cache targets data to produce errors to be exploited by Side-Channel (DFA, DEMA).
Instructions sequences are also fetched to faster cache memories that can be targeted by fault injection.
However, it is assumed that it is harder to put into practice attacks on the instruction cache than ones on the data cache:
for instance, Chen et al. say that ``timing attacks on the RSA algorithm which exploit the instruction path of a cipher are mostly proof-of-concept, and it is harder to put them into practice than D-Cache timing attacks'' in a recent (2013) paper~\cite{Chen:2013:ITI:2397723.2397914}.

\medskip
In this paper we focus on fault attack targeting the instruction cache and demonstrate their feasibility in practice.
Previous works~\cite{AB:FDTC09,B3P2:HOST10} already reported that arithmetic, logical, and branching instructions are not subject to errors in the context of global (voltage glitches) fault injection.
This fact was ascribed to the low capacitance design of CPU registers and to the presence of the instruction buffer between the CPU and the memory, which cuts down the capacitive load of the path to the instruction cache and the program memory~\cite{DBLP:series/isc/BarenghiBBPP12}.
As shown by Moro et al.~\cite{NM:HOST14} this constraint is relaxed when electromagnetic pulse injections are used, leading to single instruction replacement or skip (on 32-bit instructions) for a single pulse, or double faults when 16-bit instructions are used.
These previous results allow us to perform our experiments using only 32-bit arithmetic and logical operations (e.g., \texttt{add.w}).
This helps us to highlight and isolate the faulty behavior of the instruction read and caching mechanism from the faults occurring on the data path.

%Charts : x axis -> timing, y axis -> portion of code (c.f. papier FDTC 2014).

%*************************************
%	HOW TO ATTACK CACHE MEMORIES
%*************************************
\section{How to Attack Cache Memories}
\label{sec:cacheAttack}

\subsection{Cache Mechanisms on the Cortex-M Core}

The Cortex-M4 processor provides a three-stage pipeline Harvard architecture.
Its Flash memory interface accelerates code execution with a system of instructions prefetch buffer and cache lines.
The Cortex-M4 manages the Advanced High-performance Bus (AHB)
%\footnote{Advanced Microcontroller Bus Architecture version 2 published by ARM Ltd.}
instruction code bus and the data code bus.
This architecture provides 64 cache lines of 128 bits for instructions and 8 cache lines of 128 bits for the data.
In addition to the cache memory, a 4$\times$32 bits fast instruction buffer is in place between the CPU and the program memory.
All instructions being fetched by the CPU are preloaded into this instruction buffer.
This feature is present in most modern embedded processors.
It allows the CPU to perform aggressive instruction sequence reordering before the execution~\cite{DBLP:series/isc/BarenghiBBPP12}.
In the case of our target, instructions are always fed to the CPU from the instruction buffer, even when the instruction cache is enabled and instruction prefetch is disabled.
When the CPU finds the corresponding address of the instruction being fetched in the cache table then no wait state are needed to begin a new pipelined execution sequence.
When the CPU does not find the corresponding address then there is a cache miss and a couple of wait states are needed to refill the instruction buffer with the right program memory line.
In our case the CPU runs at 168MHz and 6 CPU cycles are needed to preload a 128 bits line of the Flash memory into the instruction buffer, and to update the cache memory according to the LRU policy.
When performing EMFI, we can expect various exploitable behaviors.
For example, if the instruction buffer update is stalled after the program pointer is updated, we can expect that four 32-bit instructions or six to height Thumb-2 (16-bit) instructions are faulted or skipped.
Another possibility, if previous instructions were already present in the instruction buffer, is that the fault injections could lead to a 128 bits cache line replacement.

%***************************************************************************************
\subsection{Experimental settings and methodology}

%In this section, we describe how we maximized our fault injection success rate by characterizing the EMFI.
%For that, we attacked three assembly test codes with three different target configurations.
\subsubsection{Experimental Setup}
\label{sec:EMplatform}

\begin{center}
\begin{figure}[!ht]
   \centering
   %\vspace*{-0.5cm}
   \includegraphics[width=\columnwidth]{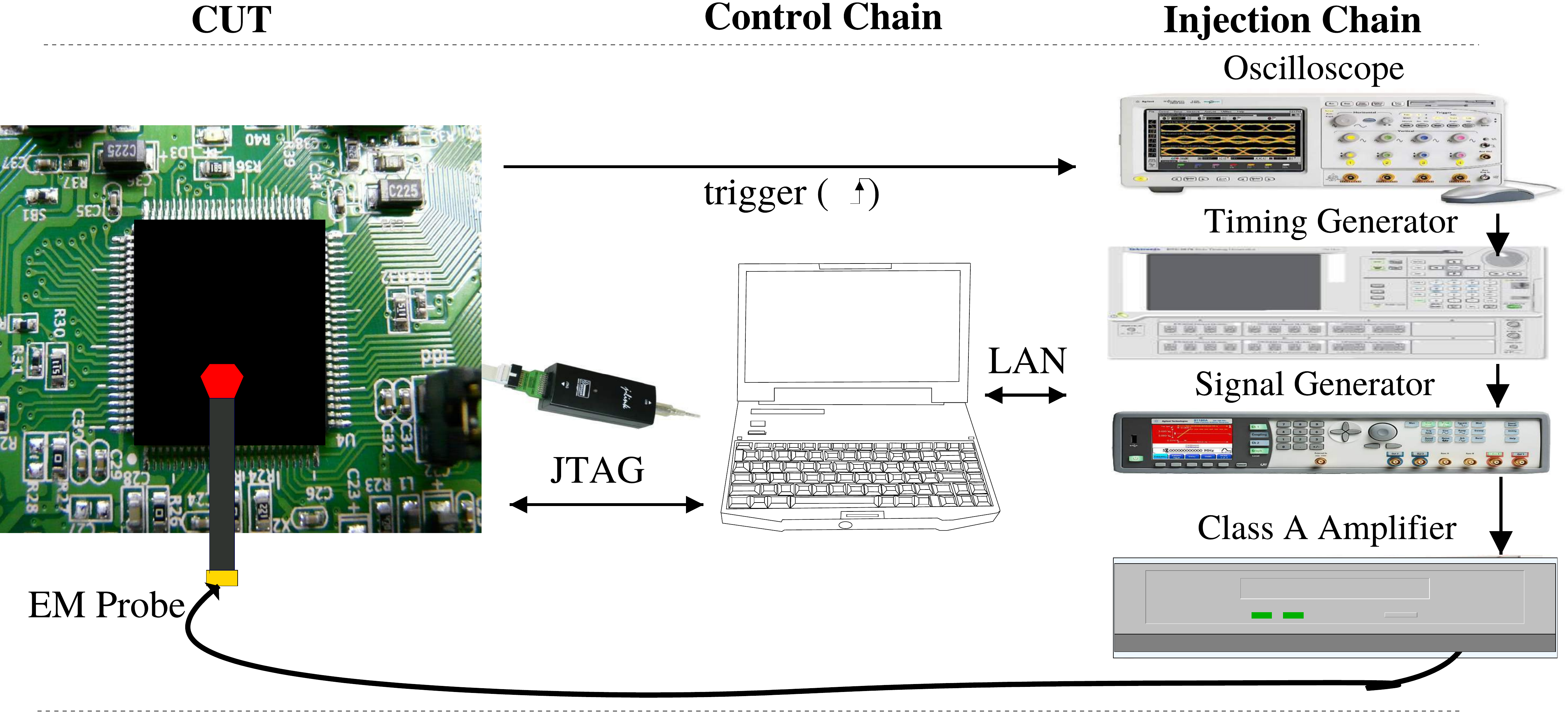}
   \caption{EM Fault Injection Platform Diagram}
   \label{fig:platform}
   \vspace*{-0.5cm}
\end{figure}
\end{center}

Figure ~\ref{fig:platform} shows an architecture view of the electromagnetic injection analysis platform used in our experiments.
It is based on a signal generator able to generate 1.5ns pulse width signal amplified by a broadband class A amplifier (400MHz, 300Watt max),
and includes an EM probe as described in~\cite{MaurineEMCE14}.
It has to be noticed the high accuracy of the signal generation (5ps of jitter) allowing us to control in a reproducible manner the fault injection.
An oscilloscope and a data timing generator are also used to control the delay before the injection.

All experiments have been performed at a settled spatial location:
a fixed position and a fixed angle orientation.
The spatial location has been determined in black-box setting (without decapsulation) in order to spot the location that maximizes the fault injection success rate.
A JTAG probe is used to dump internal register and memory content after injection.
All the necessary programs to control the platform and generate fault report have been developed, and the process is fully automatized.

 %~\ref{fig:pulse2O} .

%\begin{itemize}
%\item Setup: Electromagnetic injection platform
%\item STM32 $\rightarrow$ Oscillo $\rightarrow$ Delay generator $\rightarrow$ Pulse generator $\rightarrow$ Amp $\rightarrow$ Probe $\rightarrow$ STM32.
%\item Probe selection: Contact, near field, far field.
%\item Target: ARM Cortex-M4 based microcontroller, the STM32F407VGT6.
%  No hardware modification has been made (decapsulation).
%  Highly configurable: it is possible to enable or disable features such as prefetching, %instruction cache, data cache.
%  It is also possible to adapt the CPU core clock at will.
%  Different power mode (low 2.9-3.3V ? high 5V ?) \todo
%\end{itemize}

%* * * * * * * * * * * * * * * * * * * * * * * * * * * * * * * * * * * * * * * * * * * *
\subsubsection{Electromagnetic Fault Injection Parameters Selection}

To boost our attack control, we looked into several spatial, temporal, and energy configurations.

The spatial parameter selection is a manual step.
We determined that an optimal configuration combines two criteria:
the chip response to fault and the induced fault locality.
Therefore, we selected the spot that maximize both and obtained more than 90\% fault response on a single CPU register.

We used the configurable delay generator interposed between the oscilloscope and the pulse generator in order to cover a broad time spectrum of injection.
We observed that the delay value inserted determines which CPU register is impacted, with respect to the processed instruction sequence.
Thus, we can select the temporal parameter setting according to the chosen CPU register to be faulted.

We observed that the energy parameter selection is largely dependent on the spatial configuration.
Indeed, covering a wide range of energy level from -5.0 to +9.0 [dBm] has only a small impact on the injection success rate.
However, we found that the more power we inject, the more diffuse is the fault, implying a worse fault locality.
As a consequence, more than one CPU registers are disrupted by the attack.
We also noticed that the power range that maximizes cache fault success rate depends on the D-Cache and I-Cache configuration.
Indeed when both were enabled, it was possible to reach the best success rate at lower pulse energy (-1.7dBm) than when one or both caches were disabled (+4.5dBm).

This can be ascribed to the local voltage drop due to the load of cache memory inside the chip that increases the propagation delays.
To sum up, we fixed the spatial and energy settings to values maximizing the precision of the induced faults.
Then, we played with the temporal parameter, and performed EMFIs on different test codes.

For our experiment, 2 weeks were necessary to develop the software needed to control the platform and to automate the fault analysis. Then only few days were necessary to tune the spatial, temporal and energy configuration of the platform.
Finally, a few additional days were necessary to perform all the experiments.
%* * * * * * * * * * * * * * * * * * * * * * * * * * * * * * * * * * * * * * * * * * * *
\subsubsection{Test Codes}

\begin{wrapfigure}{r}{0.25\linewidth}
\hspace*{0.05\linewidth}
\begin{minipage}{0.95\linewidth}
\vspace*{-5mm}
\begin{lstlisting}[caption={\small \texttt{ADD Sequence} Assembly Code}, label=codeSeq, numbers=left]
add.w	r2, r2, #1
add.w	r4, r4, #1
add.w	r5, r5, #1
add.w	r6, r6, #1
add.w	r7, r7, #1
add.w	r8, r8, #1
add.w	r9, r9, #1
add.w	r10, r10, #1
add.w	r11, r11, #1
add.w	r4, r4, #5
\end{lstlisting}
\end{minipage}
\vspace*{-8mm}
\end{wrapfigure}
In order to highlight impact of induced fault on the control flow,
we build different non-idempotent assembly test code sequences, highlighting proper instruction executions.

For instance, we used a sequence of ten consecutive \texttt{ADD	Rd, Rs, \#imm} instructions.
Each of the ten instructions uses a different destination register \texttt{Rd} to store the result of the addition.

%The \texttt{ISB} (Instruction Synchronization Barrier) flushes the pipeline. After the \texttt{ISB} instruction has been completed, all following instructions are fetched from the cache or memory again.
%

%***************************************************************************************
%	PRACTICAL FAULT INJECTION ON THE CACHE
%***************************************************************************************
\section{Practical Fault Injection on the Cache}
\label{sec:cacheEMFI}

%***************************************************************************************
\subsection{Target Configurations}

We performed different fault injection campaigns according to three possible settings combinations.
For each test code, including the one depicted in Listing \ref{codeSeq}, we disabled the prefetch mechanism.
The goal is twofold: avoid faulting this step during our experiment
and increase our control over the moment when the cache loads happen.
Considering that the instruction cache and the data cache can be enabled or disabled,
there is four possible setting configurations.

\begin{table}[!ht]
\begin{center}
\begin{tabular}{cclc}
\toprule
I-Cache & D-Cache & Settings         & Practically applied \\
\midrule
0       & 0       & All off          & \Checkmark          \\
0       & 1       & D-cache only     & \XSolidBrush        \\
1       & 0       & I-cache only     & \Checkmark          \\
1       & 1       & All on (default) & \Checkmark          \\
\bottomrule
\end{tabular}
\end{center}
\caption{Settings Configuration Sets}
\label{tab:config}
\vspace*{-0.2cm}
\end{table}

As shown in Table \ref{tab:config}, we did not consider the case where only the data cache is activated as our work focuses on faulting instructions, and more generally on disrupting the control flow.
We did the first data acquisition using the default settings: both the instruction cache and the data cache were activated.
We then disabled the data cache for the second acquisition,
and finally we disabled both the data and instruction caches for the third and last data acquisition.

%***************************************************************************************
\subsection{Experimental Protocol}

Our experimental protocol consists in the following steps:

\begin{enumerate}
\item A golden run that will serve as reference:
	\begin{itemize}
  \item backup of the initial states,
  \item backup of the reference states and of the functional output.
	\end{itemize}
\item Performing EMFI:
	\begin{itemize}
  \item targeting a single instruction,
  \item retrieving output states.
	\end{itemize}
\item Classify the obtained behavior to extract fault models:
 	\begin{itemize}
  \item comparison and fault model extraction,
  \item behavioral analysis.
	\end{itemize}
\end{enumerate}

In anticipation of future fault injections, we perform a golden run (fault-free) for every test code at the first step.
We monitor the status of all general purpose and special registers and we retrieve the functional output.
We record every initial states (post-initialization values) and every reference state (post-execution values).
We also save the functional output of the targeted function if any.

In the second phase, we perform fault injections by precisely targeting a single instruction in the attacked sequence.
Accuracy is achieved by tuning the delay and power parameters:
for each temporal points from the trigger signal (steps of 1ns), the test program is executed multiple (500) times while decreasing the power by step of 0.5dBm
starting from 0dBm when I-Cache and D-Cache are activated, and from 4dBm in others cases.
After each pulse, we retrieve the output state of the function and the post-execution state of every general purpose and special registers.
The temporal point that provides high fault occurrence rate on destination register of the aforementioned instruction is used as marker to predict temporal points candidates where future instruction read and caching are likely to be performed.

The pulse delay is then tuned accordingly and only a small amount of temporal points are targeted.
The third step aims at classifying observed behavior by comparing faulted output states to the reference ones.
Based on the analysis of the observed differences, we can extract the fault model that caused the deviation from the expected state.

%***************************************************************************************
%  RESULT ANALYSIS
%***************************************************************************************
\section{Results \& Analysis}
\label{sec:results}

Our experiments revealed that the fault type we obtained by targeting the instruction read and caching mechanism is independent from the type of instruction (arithmetic-logical, branch and load/store instructions).
Previous works~\cite{AB:FDTC09,B3P2:HOST10} already reported that arithmetic, logical, and branching instructions are not subject to errors in the context of global (voltage glitches) fault injection.
This fact was ascribed to the low capacitance design of CPU registers and to the presence of the instruction buffer between the CPU and the memory, which cuts down the capacitive load of the path to the program memory.
As shown by Moro et al.~\cite{NM:HOST14} this constraint is relaxed when electromagnetic pulse injections are used, leading to single instruction replacement or skip (on 32-bit instructions) for a single pulse, or double faults when 16-bit instructions are used.

These previous results allowed us to perform our experiments using only 32-bit arithmetic and logical operations (e.g., \texttt{add.w}).
This helps us to highlight and isolate the faulty behavior of the instruction read and caching mechanism from the faults occurring on the data path.

\begin{table}[!ht]
\begin{center}
\begin{tabular}{|c|c|c|c|c|c|c|c|c|c|c|}
\hline
\#	&	\multicolumn{10}{c|}{$\longrightarrow$ Instruction flow $\longrightarrow$}  	 	\\ \cline{2-11}
NOP	&$i_1$	& $i_2$ 	& $i_3$ 	& $i_4$ 	& $i_5$ 	& $i_6$	& $i_7$	& $i_8$ & $i_9$	& $i_{10}$  \\ \hline\hline
3 	& \cref{}& \bref	& \bref	& \bref	& \bref 	& \bxef	& \bxef	& \bxef	& \bxef	& \cref{} 	\\ \hline
2	& \cref{}& \bref	& \bref	& \bref	& \bref 	& \bxef	& \bxef	& \bxef	& \bxef	& \cref{}	\\ \hline
1	& \bref	& \bref	& \bref	& \bref 	& \bxef	& \bxef	& \bxef	& \bxef	& \cref{}	& \cref{} \\ \hline
\end{tabular}

\medskip
\begin{tabular}{clclcl}
%\cline{1-3}
\cref{\ \ }	& Normal & \bref	& replayed & \bxef 	& skipped \\ %\hline
\end{tabular}
\end{center}
\caption{Impact of NOP insertion}
\label{tab:nop}
\end{table}

We used three assembly code sequences based on the same assembly macro.
What distinguishes each three of them is the memory alignment of instructions, which is determined by the insertion of one to three \texttt{nop} (No OPeration) instructions in the cache before the instruction line of the trigger signal.
As depicted in Table \ref{tab:nop}, the alignment of instructions does just shift the effect of the injection.

%***************************************************************************************
\subsection{Observed Behavior}

\begin{figure*}
  \centering
  \begin{subfigure}[t]{0.3\textwidth}
    \centering
    \includegraphics[width=\linewidth,height=2.4cm]{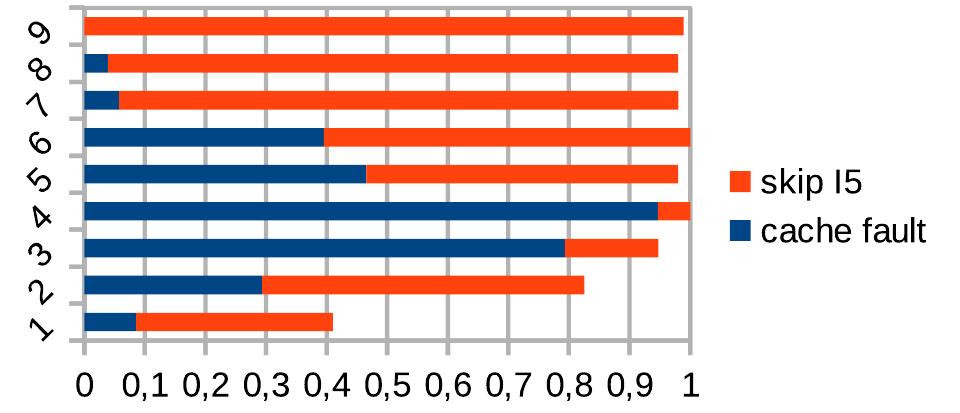}\\[-2mm]
    \caption{{\smaller D-Cache enabled, I-Cache enabled, 1 \texttt{nop}.}\label{fig:charac:1ndi}}
  \end{subfigure}
  ~
  \begin{subfigure}[t]{0.3\textwidth}
    \centering
    \includegraphics[width=\linewidth,height=2.4cm]{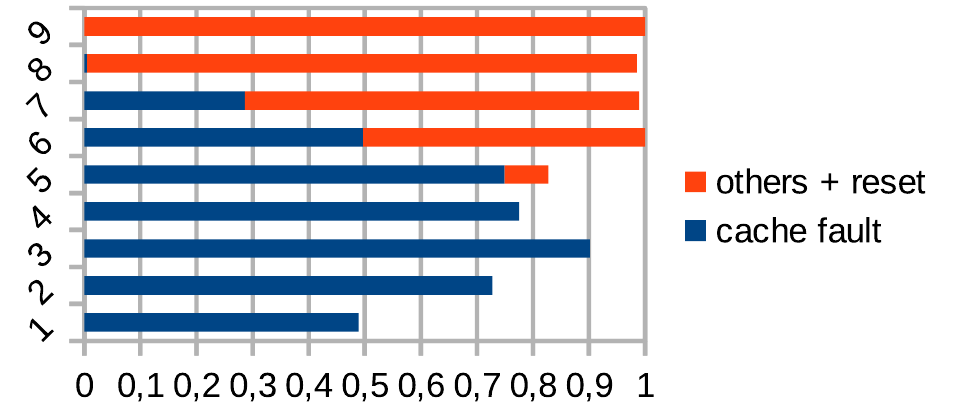}\\[-2mm]
    \caption{{\smaller D-Cache disabled, I-Cache enabled, 1 \texttt{nop}.}\label{fig:charac:1ni}}
  \end{subfigure}
  ~
  \begin{subfigure}[t]{0.3\textwidth}
    \centering
    \includegraphics[width=\linewidth,height=2.4cm]{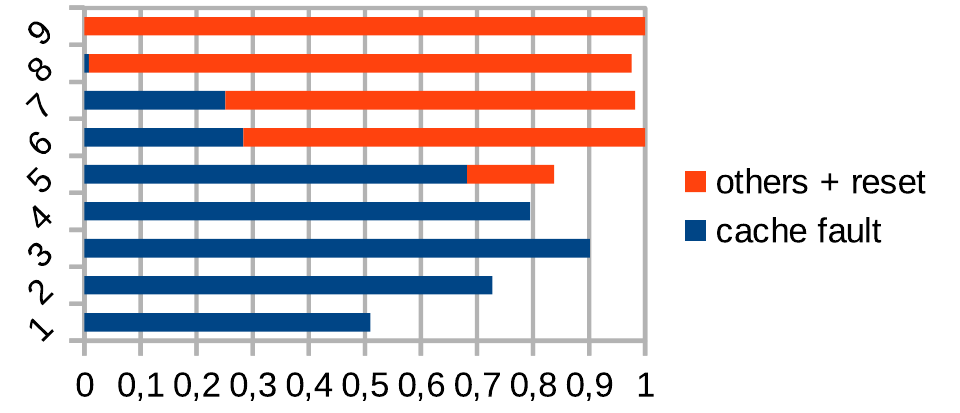}\\[-2mm]
    \caption{{\smaller D-Cache disabled, I-Cache disabled, 1 \texttt{nop}.}\label{fig:charac:1n}}
  \end{subfigure}

  \vspace*{2mm}
  \begin{subfigure}[t]{0.3\textwidth}
    \centering
    \includegraphics[width=\linewidth,height=2.4cm]{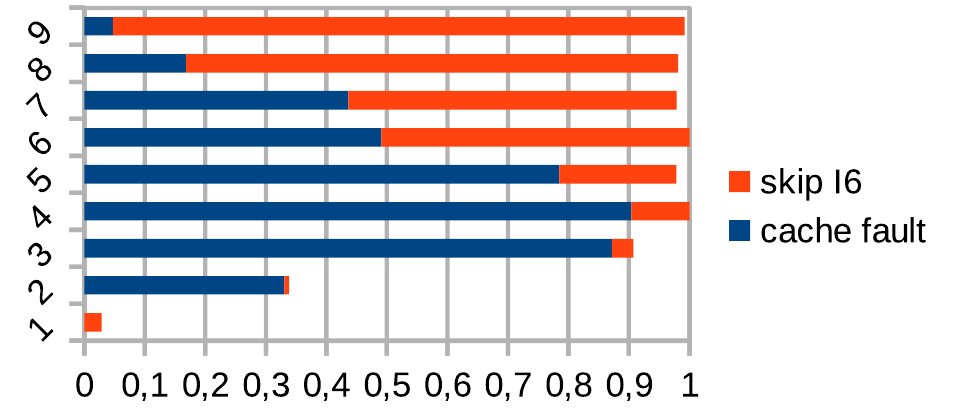}\\[-2mm]
    \caption{{\smaller D-Cache enabled, I-Cache enabled, 2 \texttt{nop}.}\label{fig:charac:2ndi}}
  \end{subfigure}
  ~
  \begin{subfigure}[t]{0.3\textwidth}
    \centering
    \includegraphics[width=\linewidth,height=2.4cm]{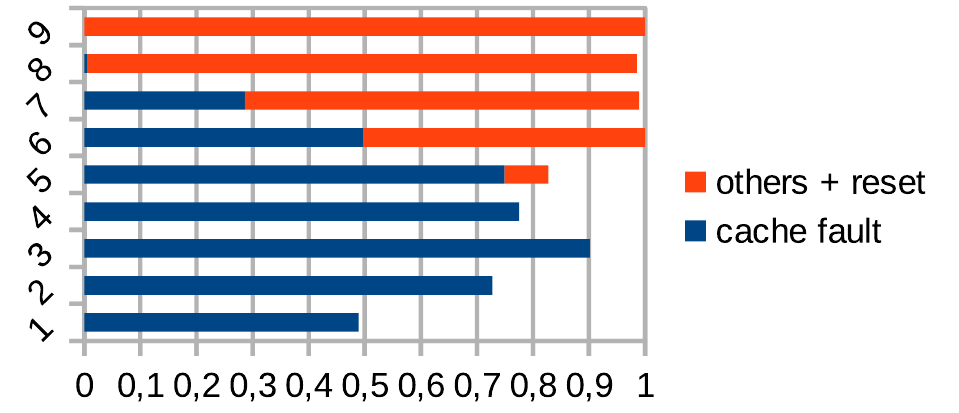}\\[-2mm]
    \caption{{\smaller D-Cache disabled, I-Cache enabled, 2 \texttt{nop}.}\label{fig:charac:2ni}}
  \end{subfigure}
  ~
  \begin{subfigure}[t]{0.3\textwidth}
    \centering
    \includegraphics[width=\linewidth,height=2.4cm]{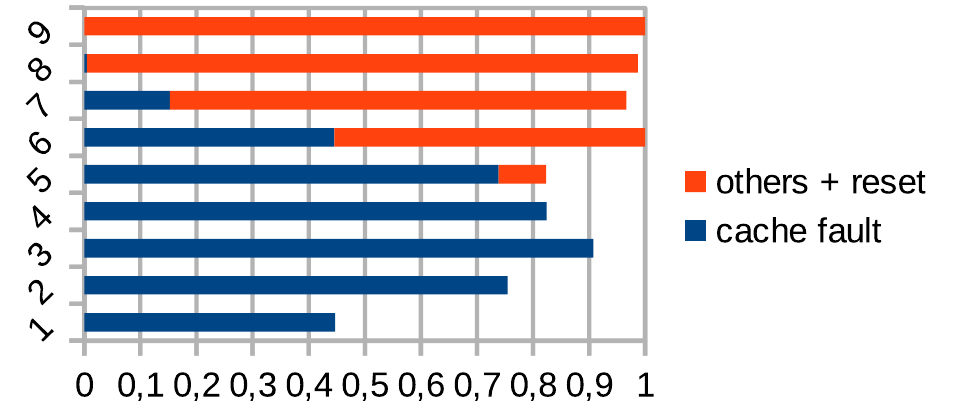}\\[-2mm]
    \caption{{\smaller D-Cache disabled, I-Cache disabled, 2 \texttt{nop}.}\label{fig:charac:2n}}
  \end{subfigure}

  \vspace*{2mm}
  \begin{subfigure}[t]{0.3\textwidth}
    \centering
    \includegraphics[width=\linewidth,height=2.4cm]{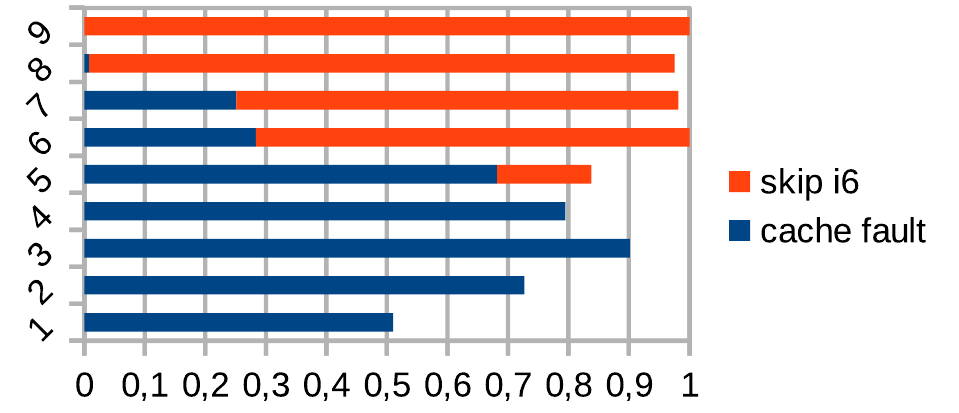}\\[-2mm]
    \caption{{\smaller D-Cache enabled, I-Cache enabled, 3 \texttt{nop}.}\label{fig:charac:3ndi}}
  \end{subfigure}
  ~
  \begin{subfigure}[t]{0.3\textwidth}
    \centering
    \includegraphics[width=\linewidth,height=2.4cm]{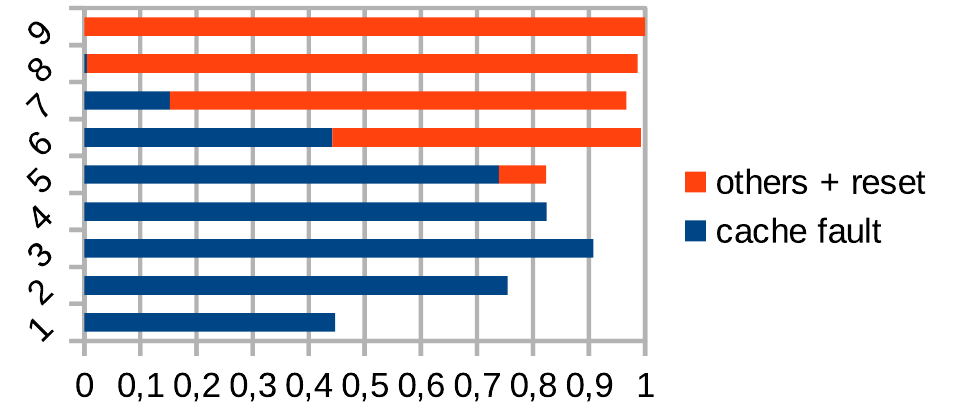}\\[-2mm]
    \caption{{\smaller D-Cache disabled, I-Cache enabled, 3 \texttt{nop}.}\label{fig:charac:3ni}}
  \end{subfigure}
  ~
  \begin{subfigure}[t]{0.3\textwidth}
    \centering
    \includegraphics[width=\linewidth,height=2.4cm]{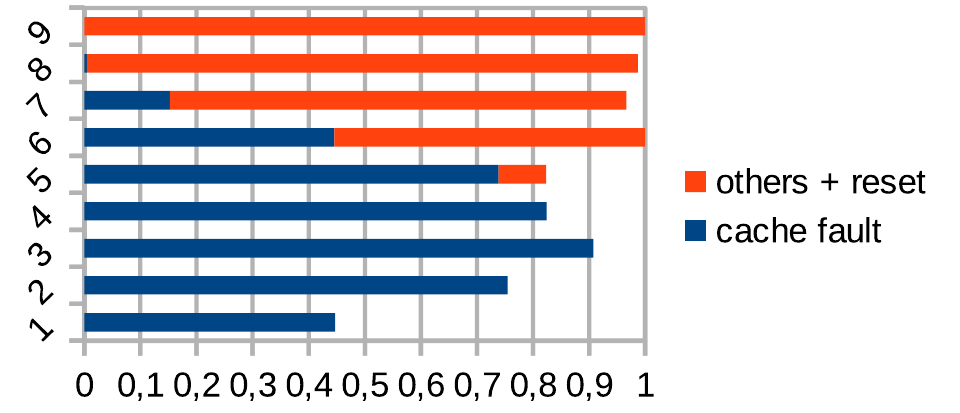}\\[-2mm]
    \caption{{\smaller D-Cache disabled, I-Cache disabled, 3 \texttt{nop}.}\label{fig:charac:3n}}
  \end{subfigure}

  \caption{\label{fig:charac} Fault model probability depending on injected power.}
  \vspace*{-3mm}
\end{figure*}

Figure \ref{fig:charac} describes the fault model occurrence probability observed during our experiment with respect to the power level of the fault injection, according to the protocol defined in Section \ref{sec:cacheAttack}.
For every figures, the power of the injection is represented on the abscissa and starts from a reference value denoted as 0.
The step between each tenth of the abscissa takes the constant value of 0.5dBm.
The ordinate axis transcribes the index of the targeted instruction in the test code.

The configuration of the D-Cache and the I-Cache varies according to the columns: For figures $\{a,d,g\}$ on the 1$^\text{st}$ column, both D-Cache and I-Cache are enabled.
For figures $\{b,eh\}$ on the 2$^\text{nd}$ column, only the I-Cache is activated and for figures $\{c,f,i\}$  on the 3$^\text{rd}$ column, all caches are disabled.

Finally, the code alignment varies according to the rows: For figures $\{a,b,c\}$ on the 1$^\text{st}$ row, there is only one \texttt{nop} offset, two for figures $\{d,e,f\}$ and three for figures $\{g,h,i\}$ on the 3$^\text{rd}$ row.

%****************
\bigskip
For a given sequence of 32-bit instructions, when targeting the $n^\text{th}$ instruction $i_n$ we observed that:
\begin{itemize}
\item instructions \{$i_n$, $i_{n+1}$, $i_{n+2}$, $i_{n+3}$\} are skipped,
\item \{$i_{n-4}$, $i_{n-3}$, $i_{n-2}$, $i_{n-1}$\} are replayed instead,
\item the execution continues at instruction $i_{n+4}$.
\end{itemize}

\begin{table}[!ht]
\vspace*{3mm}
\begin{center}
\begin{tabular}{r|c|c|c|c|c|}
%\cline{1-1}\cline{3-3}\cline{5-5}
\cline{2-2}\cline{4-4}\cline{6-6}
&Instruction		& 		& Cache 	& & Resulting 		\\
&Flow				& 		& Read	& & flow				\\ \cline{2-2}\cline{4-4}\cline{6-6}
&...				& \mrow{2}{$\rightarrow$}	& \mrow{2}{$Read_{128}$}& \mrow{2}{$\rightarrow$} & ...	\\
&$i_{n-5}$		& 		& 		& & $i_{n-5}$ 		\\ \cline{2-2}\cline{4-4}\cline{6-6}
&$i_{n-4}$		& 		& 		& & $i_{n-4}$		\\
&$i_{n-3}$		& \mrow{2}{$\rightarrow$}& \mrow{2}{$Read_{128}$}	& \mrow{2}{$\rightarrow$}& $i_{n-3}$	\\
&$i_{n-2}$		& 		& 		& & $i_{n-2}$		\\
&$i_{n-1}$		& 		& 		& & $i_{n-1}$		\\ \cline{2-2}\cline{4-4}\cline{6-6}
$\leadsto$&\crst{$i_n$}	& 		& \bxef	& & \cini{$i_{n-4}$}	\\
&\crst{$i_{n+1}$}& \mrow{2}{$\varnothing$}& \bxef & \mrow{2}{$\circlearrowright$} & \cini{$i_{n-3}$}	\\
&\crst{$i_{n+2}$}	& 		& \bxef	& & \cini{$i_{n-2}$}	\\
&\crst{$i_{n+3}$}	& 		& \bxef	& & \cini{$i_{n-1}$}	\\ \cline{2-2}\cline{4-4}\cline{6-6}
&$i_{n+4}$		& \mrow{3}{$\rightarrow$}& \mrow{3}{$Read_{128}$}	& \mrow{3}{$\rightarrow$} & $i_{n+4}$\\
&$i_{n+5}$		& 		& 		& & $i_{n+5}$		\\
&...				& 		& 		& & ...				\\ %\cline{2-2}\cline{4-4}\cline{6-6}
\end{tabular}

\medskip
\begin{tabular}{clcl}
%\cline{1-3}
 \cini{}	& replayed & \crst{} 	& skipped \\ %\hline
\end{tabular}
\end{center}
\caption{Faulting Cache Read}
\label{tab:cacheRead}
%\vspace*{-2mm}
%\vspace*{-2cm}
\end{table}

\vspace{-1mm}
%\medskip
As shown in Table \ref{tab:cacheRead}, this behavior corresponds to preventing the update of the instruction buffer, also denoted as prefetch queue (PFQ).
Indeed, at each cache read operation, 128 bits (four 32-bit instructions or up to eight 16-bit instructions) of data are prefetched.
%Zak: c'est 6 cycles (5WS)
A new cache read is performed every 6 CPU clock cycles when the prefetch mechanism is disabled.
At the time of the instruction read just before the fault injection, denoted $t_{-1}$,
the sequence of four instructions which is present in the instruction buffer is
$Cache(t_{-1}) =$ \{$i_{n-4}$, $i_{n-3}$, $i_{n-2}$, $i_{n-1}$\}.
At $t_0$, the next sequence of four instructions is supposed to be loaded in the instruction buffer.
However the fault somehow disrupt the memory read operation from the Flash to the instruction buffer.
Therefore, $Cache(t_{0}) = Cache(t_{-1})$, which results in replaying the four previous instructions: \{$i_{n-4}$, $i_{n-3}$, $i_{n-2}$, $i_{n-1}$\}.

However, the program counter continues to increment normally so that at the next instruction read we have
$Cache(t_{1}) =$ \{$i_{n+4}$, $i_{n+5}$, $i_{n+6}$, $i_{n+7}$\}.
Consequently, the four instructions \{$i_n$, $i_{n+1}$, $i_{n+2}$, $i_{n+3}$\} are never loaded into the instruction buffer, and are thus simply skipped.

%***************************************************************************************
%  DISCUSSION
%***************************************************************************************
\section{Discussions}
\label{sec:discuss}

\subsection{Possible Use Case for Standard Attacks}

Our fault model has several advantages over the ones existing in the literature.
First, it is reproducible, in the sense that for a given input we can expect that the faulted output will be the same in up to 96\% of the cases.
Second, the same fault model can be repeated several times during a computation, and at very close intervals.
Such a predicable model can be very helpful to mount combined attacks using SCA and FA.
Combined SCA and FA attacks are powerful because they allow to break SCA and FA-protected implementations of SP-Networks~\cite{RocheLK11}.
Yet, they are considered difficult because of the necessity for the fault model to be reproducible at wish.
Indeed, in order to make use of a statistical analysis such as a SCA on a faulty computation, one must be able repeat the very same fault several times.
In our case, such an attack would be straightforward.

\begin{table}[!ht]
\begin{center}
\begin{tabular}{cllc}
\toprule
Id & Classical Attack     & Effect on CF  & Apply      \\
\midrule
1  & DFA on AES           & Replay        & \Checkmark \\
2  & BellCoRe on CRT-RSA  & 4B difference & \Checkmark \\
3  & Privilege escalation & FreeRTOS      & \Checkmark \\
4  & Nullify XOR          & Replay        & \Checkmark \\
\bottomrule
\end{tabular}
\end{center}
\caption{Faulting Cache Reachable Attack on Non-secured Implementations}
\label{tab:reach}
\vspace*{-1mm}
\end{table}

As shown in Table \ref{tab:reach}, attacking the cache makes it possible to mount classical attacks.
For example, the one by Dehbaoui et al.~\cite{Dehbaoui2013Cosade} is achievable with our fault model:
we can avoid the AES round counter increment one or several times in order to increase the leakage of manipulated round keys.
Thereby, we can perform a differential fault analysis (DFA) on the AES.

%\begin{table}[!ht]
%\begin{center}
%\vspace*{3mm}
%\begin{tabular}{cllc}
%\toprule
%Id & Attack & Effect on CF                & Apply      \\
%\midrule
%1  & DFA    & Avoiding CTM mechanism      & \Checkmark \\
%2  & FA     & Nullifying Instr. dupl.     & \Checkmark \\
%3  & FA     & Avoiding HW CTM activation* & \Checkmark \\
%\bottomrule
%\end{tabular}
%\end{center}
%\caption{Faulting Cache Attack on Fault and SCA Countermeasures}
%\label{tab:reachCTM}
%*{Not enough time to reset the memory for instance}
%\vspace*{-0.5cm}
%\end{table}

\subsection{Simplifying Existing Attacks and Deploying New Attacks}

Here we give a few insights of potential usage for our fault model.
We believe some render practical existing but rather theoretical attacks (or at least simplify them),
while others open new attack paths.
All of these would be interesting to study in further work.

Infective countermeasures are sometimes chosen based on the assumption that skipping a branching instruction is easier to do than zeroing the value of a big number,
for instance in CRT-RSA countermeasures against the BellCoRe attack as explained by Rauzy and Guilley in their FDTC 2014 paper~\cite{6976633}.
Our fault model shows that it actually is realistic to mount such high-level zeroing attacks.

In their paper at CCS 2014~\cite{DBLP:journals/iacr/BartheDFGZ14}, Barthe et al. expose new fault attacks on the CRT-RSA cryptosystem.
The idea of the proposed fault attacks is to zeroize the most significant bits of one of the intermediate exponentiations.
The required fault injection to achieve their new attack is somewhat complicated:
it requires forcing some registers at zero in addition to faulting the control flow of the modular exponentiation.
Here too, repeating our fault model enough times to entirely bypass the exponentiation (e.g., repeating the instructions which increment the loop counter) would simplify the work for the attacker.

\begin{wrapfigure}{r}{0.2\linewidth}
\hspace*{0.05\linewidth}
\begin{minipage}{0.95\linewidth}
\vspace*{-4mm}
\begin{lstlisting}[caption=Masking,numbers=left,label=lst_idea_mask]
mov  r5, #m
ldr  r0, [r5]
eor  r1, r1, r0
ldr  r0, [r5, #1]
eor  r2, r2, r0
ldr  r0, [r5, #2]
eor  r3, r3, r0
ldr  r0, [r5, #3]
eor  r4, r4, r0
\end{lstlisting}
\end{minipage}
\vspace*{-8mm}
\end{wrapfigure}
Now, consider a typical implementation of a masked block cipher (e.g., AES).
We have for instance 128 bits of state stored in registers $r_1$, $r_2$, $r_3$, and $r_4$.
The mask is stored in the memory address $m$, which is loaded into the register $r_5$.
To apply the mask, a sequence of loading and xoring instructions would be used as in Listing~\ref{lst_idea_mask}.
In that case, our fault model would allow the attacker to replay the first 4 instructions and skip the last 4, effectively unmasking the whole state and helping a side-channel attacker to break the masking countermeasure.
That of course is a best case scenario, in which the instructions are perfectly aligned with our need in the instruction buffer;
in practice it maybe possible to unmask only a part of the state, which would still benefit to a side-channel attacker.

Yet another possible use case would be to break the countermeasure against instruction skip presented by Moro et al. in their JCEN paper~\cite{MoroJCEN14}.
Their countermeasure is formally proven to protect against single-instruction skip.
However, since we skip at least 4 instructions (in the case of 32-bit instructions) in our fault model, their countermeasure would not work against the model of fault we presented in this paper.

\subsection{Conclusions and Perspectives}

We presented a practical electromagnetic fault injection (EMFI) methodology that applies to all the ARMv7-M architectures.
This method provides high precision and high reproducibility.
Indeed, the precise fault model we described occurs in up to 96\% of the cases when the EMFI settings are tuned according to our description.
The induced fault model was shown practically useful for attackers, in particular against cryptographic algorithms, including ones equipped with countermeasures against fault injections.

As future work, it would be interesting to study the feasibility of an efficient countermeasure against our fault model.
Another obvious next step is to put our EMFI into practice on more systems and study its portability onto other architectures than ARMv7-M.

%***************************************************************************************
%	BIBLIOGRAPHY
%***************************************************************************************

\bibliographystyle{unsrt}
\bibliography{./article,./fia}

\end{document}